\documentclass[final]{aipproc}

\layoutstyle{6x9}

\begin{document}

\title{New Physics constraints from optimized observables in $B\to K^* \mu^+\mu^-$ at large recoil}


\author{S\'ebastien Descotes-Genon}{
  address={Laboratoire de Physique Th\'eorique, CNRS/Univ. Paris-Sud 11 (UMR 8627)\\ 91405 Orsay Cedex, France}
}

\author{Joaquim Matias}{
  address={Universitat Aut\`onoma de Barcelona, 08193 Bellaterra, Barcelona, Spain}
}

\author{Javier Virto}{
  address={Universitat Aut\`onoma de Barcelona, 08193 Bellaterra, Barcelona, Spain}
  ,altaddress={Speaker} 
}

\begin{abstract}
$B\to K^* \mu^+\mu^-$ angular observables have become a key ingredient in global model-independent analyses of $b\to s$ transitions. However, as experimental precision improves, the use of theoretically clean quantities becomes a crucial issue. Global analyses that use \emph{clean} observables integrated in small bins are already a reality, opening up a new chapter in our quest for New Physics.
\end{abstract}

\maketitle


\section{Status of $\ b\to s\ \ell^+\ell^-$ decays}

During the last few years, intensive theoretical work and impressive experimental results --and even more impressive prospects-- have pushed our understanding on $b\to s\ \ell^+\ell^-$ decays far beyond expectations. Among the large set of inclusive and exclusive $b\to s\ \ell^+\ell^-$ modes, a considerable attention has been put into $B\to K^{(*)}\mu^+\mu^-$, where experimental analyses are based on ${\cal O}(10^2)$ events in the case of the B-factories and CDF \cite{0904.0770,1204.3933,1108.0695} and ${\cal O}(10^3)$ events at LHCb \cite{LHCbinned}. Also, recent bounds on the $B_s\to \mu^+\mu^-$ branching ratio \cite{Bsmumu} are very close to the SM prediction, taking into account the correction from the $B_s$ width difference to branching ratio measurements at LHCb \cite{1111.4882,1204.1735+1204.1737}. These modes are very sensitive to New Physics contributing to right-handed currents (since transverse asymmetries in $B\to K^{(*)}\mu^+\mu^-$ measure indirectly the polarization of the virtual photon) and to scalar and pseudo-scalar operators (specially in the case of $B_s\to \mu^+\mu^-$).
 
 The theoretical description of $b\to s\ \ell^+\ell^-$ decays within and beyond the SM is given by the $\Delta B=-\Delta S=1$ effective Hamiltonian ${\cal H}_{\rm eff}=\sum_i C_i {\cal O}_i$ \cite{Chetyrkin:1996vx,Bobeth:1999mk}. In the SM the relevant operators are the electromagnetic dipole and semileptonic operators ${\cal O}_7$, ${\cal O}_9$ and ${\cal O}_{10}$. The 4-quark current-current ${\cal O}_{1,2}^{u,c}$ and QCD-penguin ${\cal O}_{3,4,5,6}$ operators and the chromo-magnetic dipole operator ${\cal O}_8$ are involved at higher orders in perturbation theory. Beyond the SM, non-standard operators become important, such as the chirality-flipped operators ${\cal O}_{7,9,10}'$, scalar ${\cal O}_{S^{(\prime)}}$, pseudo-scalar ${\cal O}_{P^{(\prime)}}$ and tensor ${\cal O}_{T,T5}$ operators.

 The theoretical description of the $B\to K^{*}\mu^+\mu^-$ decay becomes uncontrollable when the invariant dilepton mass $q^2$ approaches the threshold of $q\bar q$ resonance production. This happens predominantly in the vicinity of the $\psi$ and $\psi'$ $c\bar c$ states, around $q^2\sim 8-15$ GeV$^2$. The theoretical methods used to describe the regions below (low-$q^2$) and above (large-$q^2$) the vetoed range are different. The effect of a finite width of the $K^*$ including two scalar resonances has been addressed in Ref.~\cite{Becirevic:2012dp}, pointing to a non-negligible impact in some observables at low-$q^2$. However, an experimental fit to certain \emph{folded} angular distributions can decouple these effects \cite{mat}.
 
 At large recoil, the transversity amplitudes are computed in QCD factorization in the large energy limit of the $K^*$ \cite{0106067,0412400}. In this limit, symmetry relations between the seven heavy-to-light form factors allow one to express the amplitudes in terms of two soft form factors $\xi_{\|,\bot}$, distribution amplitudes and calculable hard kernels up to ${\cal O}(\alpha_s,\Lambda_{QCD}/m_b)$ \cite{Charles:1998dr} (see also \cite{Beneke:2000wa}). Soft gluon contributions at the tail of $c\bar c$ resonances in the low-$q^2$ region have been computed in Ref.~\cite{1006.4945}. Corrections of order ${\cal O}(\Lambda_{QCD}/m_b)$ are unknown, and include symmetry-breaking contributions to form factor relations and non-factorizable contributions from distribution amplitudes.

 Currently, the main uncertainties in the prediction of angular observables in $B\to K^{*}\mu^+\mu^-$ are due to unknown ${\cal O}(\Lambda_{QCD}/m_b)$ corrections and hadronic uncertainties in form factor computations from light-cone sum rules \cite{1006.4945,ffs}. The efforts to reduce these uncertainties in phenomenological applications have  led to the identification of \emph{clean} or \emph{optimized} observables, defined as ratios where most of the dependence on form factors cancels. A complete list of such observables is given by $A_T^{(2,3,4,5)}$ \cite{AT},  $A_T^{\rm (re, im)}$ \cite{1106.3283}, $P_{1,2,3}$, $P_{4,5,6}^{(\prime)}$, $M_{1,2}$ and $S_{1,2}$ \cite{1202.4266,1207.2753} at low-$q^2$ and $H_T^{(2,3,4,5)}$ \cite{1006.5013} at high-$q^2$. Experimental analyses have focused on the measurements of the branching ratio $BR$, the forward-backward asymmetry $A_{\rm FB}$, the longitudinal polarization fraction $F_L$, $A_{im}$ and $S_3$ (see \cite{0811.1214}), always integrated in a series of $q^2$ bins. CDF has measured directly the optimized observable $A_T^{(2)}\equiv P_1$ \cite{1108.0695}, while the observables $P_{1,2,3}$ can be obtained indirectly from the LHCb results (see Ref.~\cite{1207.2753} and Table \ref{JVPis}).

A wealth of model-independent combined analyses of $b\to s\gamma$ and $b\to s\ell^+\ell^-$ decays have appeared recently in the literature \cite{1207.2753,1104.3342+1202.2172,1105.0376+1111.2558,1111.1257+1206.0273,1205.1838,1206.1502}. The differences include the statistical treatment, the set of observables included in the analysis, and the NP scenarios considered. All in all, the data is compatible with the SM, as well as with the flipped-sign point $C_{7,9,10}=-C_{7,9,10}^{\rm \scriptscriptstyle SM}$. For a more thorough status review of $b\to s\ \ell^+\ell^-$ see Refs.~\cite{1208.3057,1208.3355}.

\section{Clean observables in $B\to K^* \mu^+\mu^-$ at large recoil}

Based on the symmetries of the angular distribution discussed in Ref.~\cite{1005.0571}, the minimum number of observables needed to describe the full $B\to K^{*}\mu^+\mu^-$ angular distribution can be inferred, which varies depending on whether mass and/or scalar effects are considered. A \emph{basis} of angular observables can then be identified, with the property of containing a minimum number of observables from which \emph{any} other observable can be obtained.
The basis is not unique, but there is a subset of bases with a quality feature: they contain a maximum number of \emph{clean} observables. One such bases has been constructed and studied in detail in Refs.~\cite{1202.4266,1207.2753}:
\begin{equation}
O=\Big\{\frac{d\Gamma}{dq^2},A_{\rm FB},P_1,P_2,P_3,P'_4,P'_5,P'_6,M_1,M_2,S_1,S_2\Big\}\ .
\end{equation}
All but $d\Gamma/dq^2$ and $A_{\rm FB}$ are clean observables, whereas\footnote{The scalar observables $S_{1,2}$ here should not be confused with the observables in Ref.~\cite{0811.1214}.} $S_i$ vanish in the absence of contributions from scalar operators, and $M_i$ go to zero in the limit of zero lepton masses. While the observables $M_2$ and $S_{1,2}$ are very much constrained by the $B_s\to \mu^+\mu^-$ branching ratio, the rest shows a good sensitivity to New Physics, especially $P_{1,2}$, $P'_{4,5}$ \cite{1202.4266,1207.2753}. From the latest LHCb measurements of $B\to K^{*}\mu^+\mu^-$ \cite{LHCbinned}, experimental results can be derived for $P_{1,2,3}$ integrated in the bins $[2,4.3]$, $[4.3,8.68]$ and $[1,6]$ GeV$^2$. The experimental numbers together with the theoretical predictions are displayed in Table \ref{JVPis}.

\begin{table}
\centering
\begin{tabular}{lrr}
\hline\hline
Observable & Experiment & SM prediction \\
\hline
$\langle P_1 \rangle_{[2,4.3]}$ & $-0.19 \pm 0.58$ & $-0.051 \pm 0.050$ \\
\hline
$\langle P_1 \rangle_{[4.3,8.68]}$ & $0.42 \pm 0.31$ & $-0.117\pm 0.059$ \\
\hline
$\langle P_1\rangle_{[1,6]}$ & $0.29 \pm 0.47$ & $-0.055 \pm 0.051$\\
\hline
$\langle P_2\rangle_{[2,4.3]}$ & $0.51 \pm 0.27$ & $0.232 \pm 0.069$\\
\hline
$\langle P_2\rangle_{[4.3,8.68]}$ & $-0.25 \pm 0.08$ & $-0.405\pm 0.064$ \\
\hline
$\langle P_2\rangle_{[1,6]}$ & $0.35 \pm 0.14$ & $0.084 \pm 0.066$\\
\hline
$\langle P_3\rangle_{[2,4.3]}$ & $0.08 \pm 0.35$ & $-0.004 \pm 0.024$\\
\hline
$\langle P_3\rangle_{[4.3,8.68]}$ & $-0.05 \pm 0.16$ & $-0.001\pm 0.027$ \\
\hline
$\langle P_3\rangle_{[1,6]}$ & $-0.21 \pm 0.21$ & $-0.003\pm 0.024$\\
\hline\hline
\end{tabular}
\caption{Experimental values for the clean observables $P_1$, $P_2$ and $P_3$ within different $q^2$-bins, extracted from the measurements of $S_3$, $A_{\rm im}$, $A_{\rm FB}$ and $F_L$, and their SM predictions.}
\label{JVPis}
\end{table}

\begin{figure}
\begin{minipage}{\textwidth}
\hspace{-0.5cm}
\includegraphics[height=5cm,width=7cm]{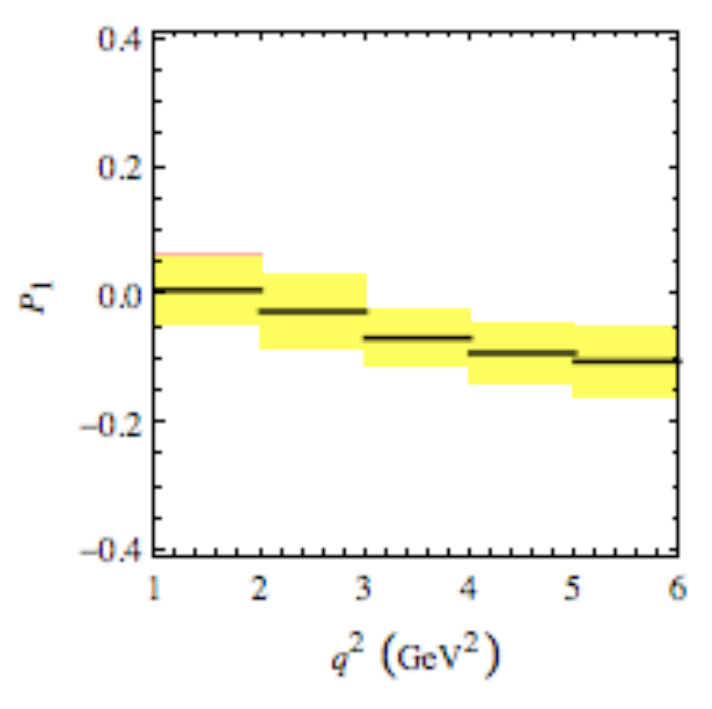}\hspace{0.5cm}
\includegraphics[height=5cm,width=7cm]{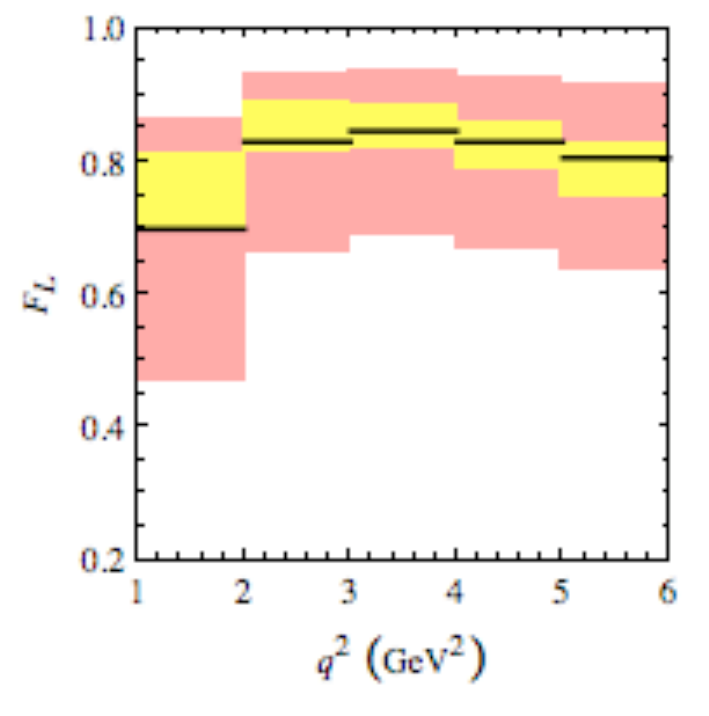}\\
\includegraphics[height=5cm,width=6.5cm]{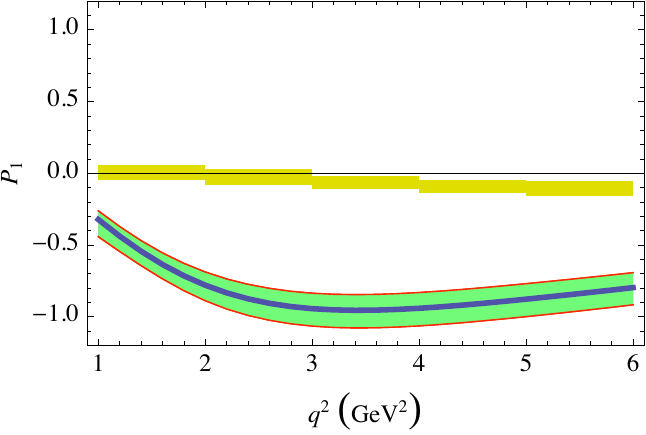}\hspace{1cm}
\includegraphics[height=5cm,width=6.5cm]{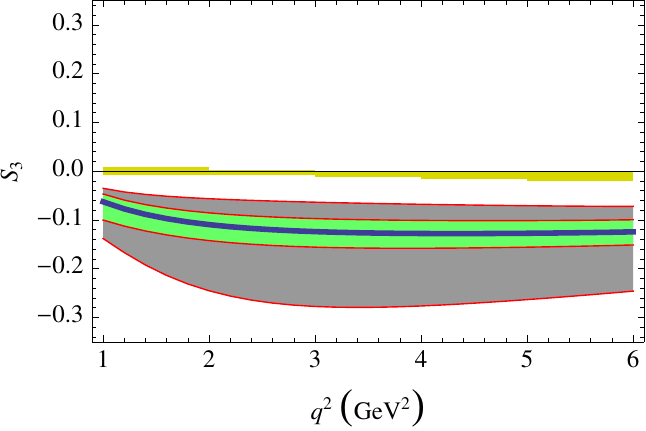}
\end{minipage}
\caption{Comparison between the observables $P_1$, $F_L$ and $S_3$ concerning their dependence on hadronic uncertainties. See the text for details.
}
\label{JVP1S3fig}
\end{figure}

The suppressed dependence on hadronic uncertainties of the clean observables $P_i$ compared to other observables can be checked directly. In the upper plots in Fig.~\ref{JVP1S3fig}, we show the SM predictions for $P_1$ and $F_L$, including all uncertainties, with the form factors taken from Ref.~\cite{ffs} (yellow) and from Ref.~\cite{1006.4945} (red) --this last reference is more conservative in the error treatment--. The conclusion is that, while $P_1$ is basically insensitive to this choice, the theoretical error in $F_L$ can vary by more than a factor of 2, with uncertainties up to a 30\%. In the lower plots in Fig.~\ref{JVP1S3fig}, a similar comparison is performed between $P_1$ and the corresponding observable $S_3$ of Ref.~\cite{0811.1214}. In this case the yellow boxes are the SM predictions, the blue curve is a NP benchmark point consistent with all other data (benchmark point `$b2$' in \cite{1207.2753}), with the green band corresponding to the total uncertainty taken the form factors in \cite{ffs} and the gray band for the form factors in \cite{1006.4945}. While the observable $S_3$ is protected from form factor uncertainties near the SM point, we can see that this is no longer true around other allowed regions in the parameter space. While this benchmark point is clearly discernible from the SM measuring $P_1$ with a 20\% error, a measurement of $S_3$ will hardly bring a definite conclusion. These examples demonstrate the importance of focusing on clean observables.

\section{Model-Independent constraints}

A combined model-independent analysis including constraints from $BR(B\to X_s\gamma)$, $S_{K^*\gamma}$, $A_I(B\to K^*\gamma)$, $BR(B\to X_s\mu^+\mu^-)$, together with binned observables in $B\to K^* \mu^+\mu^-$ at low-$q^2$ has been presented in Ref.~\cite{1207.2753}. $B\to K^* \mu^+\mu^-$ observables include the forward-backward asymmetry, $F_L$, and $P_{1,2,3}$.

\begin{figure}
\includegraphics[height=6.5cm]{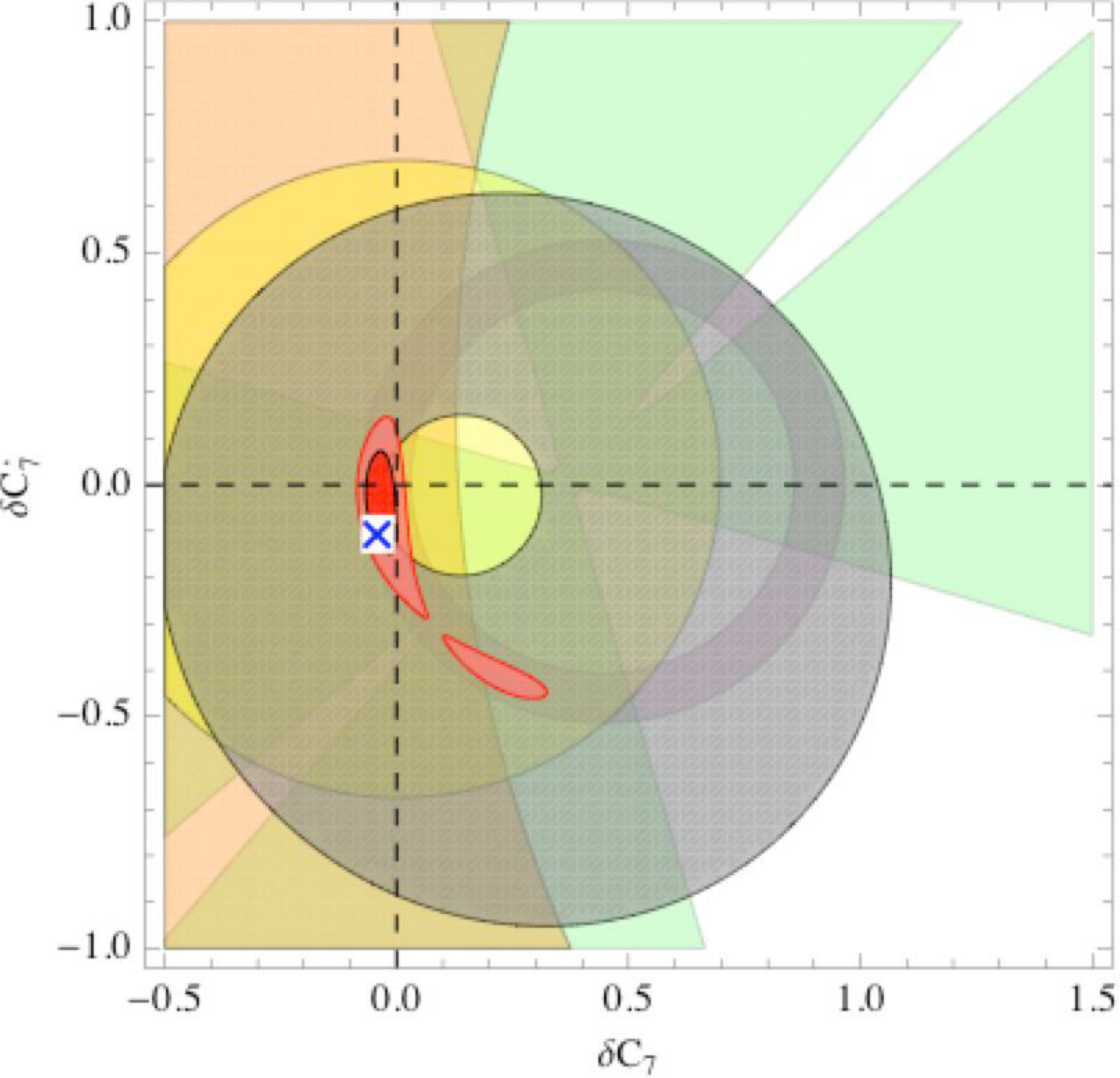}\hspace{0.7cm}
\includegraphics[height=6.5cm]{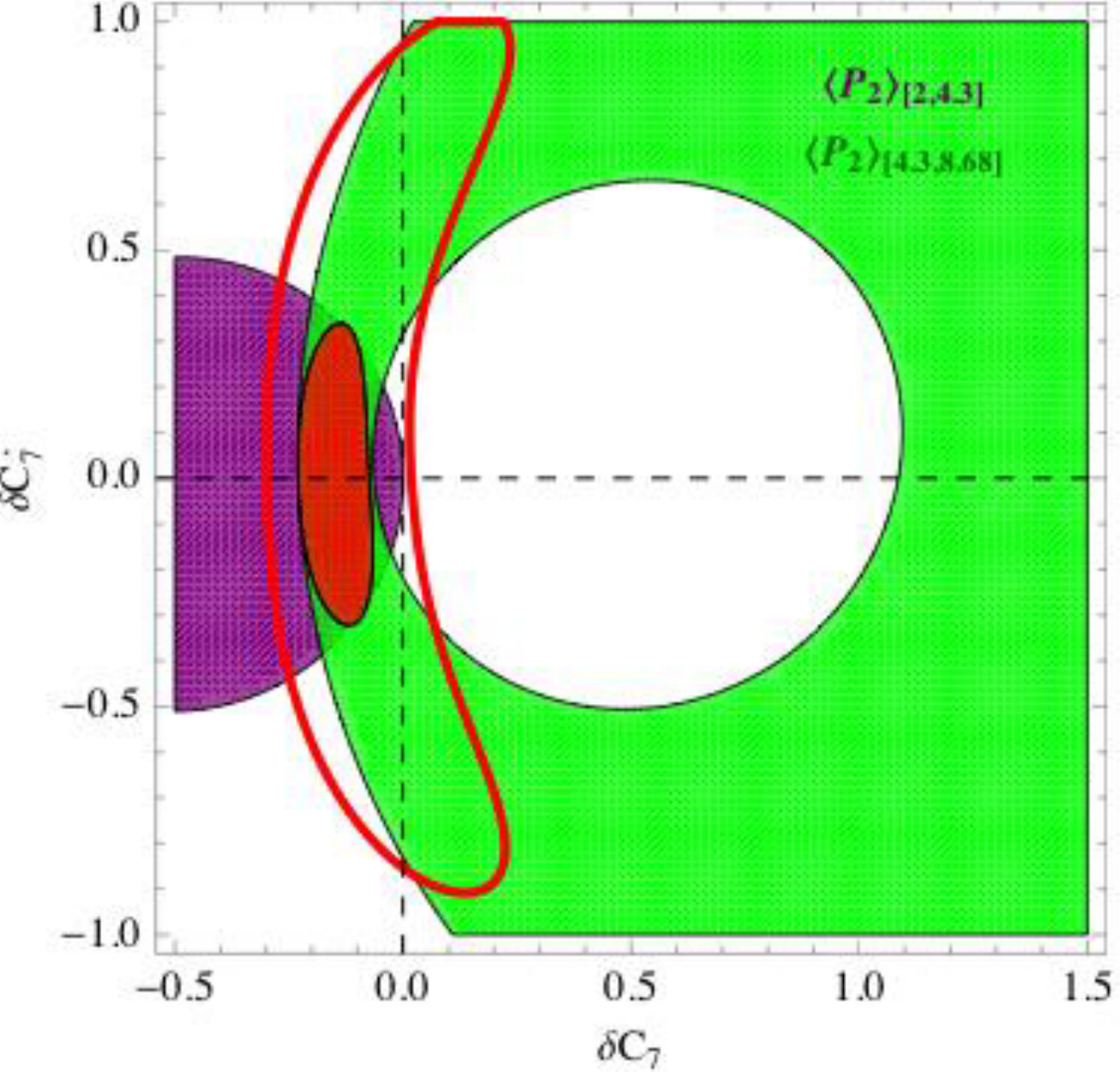}
\caption{Left: 68.3\% and 95.5\% C.L. constraints on $C_7,C_7'$ from $BR(B\to X_s\gamma)$, $S_{K^*\gamma}$, $A_I(B\to K^*\gamma)$, $BR(B\to X_s\mu^+\mu^-)$, $\langle A_{\rm FB} \rangle_{[1,6]}$ and $\langle F_L \rangle_{[1,6]}$. Right: 68.3\% and 95.5\% C.L. constraints on $C_7,C_7'$ from $\langle P_2 \rangle_{[2,4.3]}$ and $\langle P_2 \rangle_{[4.3,8.68]}$. The notation is $C_7=C_7^{\rm SM}+\delta C_7$, and similarly for $C_7'$.
}
\label{JVcons}
\end{figure}

In Fig.~\ref{JVcons} (left) we show the 68.3\% and 95.5\% C.L. combined constraints on $C_7,C_7'$ from $BR(B\to X_s\gamma)$, $S_{K^*\gamma}$, $A_I(B\to K^*\gamma)$, $BR(B\to X_s\mu^+\mu^-)$, $\langle A_{\rm FB} \rangle_{[1,6]}$ and $\langle F_L \rangle_{[1,6]}$. In the right plot of the same figure, the constraints from $\langle P_2 \rangle_{[2,4.3]}$ and $\langle P_2 \rangle_{[4.3,8.68]}$ are shown. While the experimental numbers for $\langle P_2\rangle_{\rm bin}$ must be still improved considerably (the values used do not include correlations), the constraints from $\langle P_2\rangle_{\rm bin}$ are already interesting in comparison with the combined constraints from the other observables. Both bins point towards negative $\delta C_7$. This result is not affected by form factor uncertainties.

\begin{figure}
\includegraphics[height=8cm]{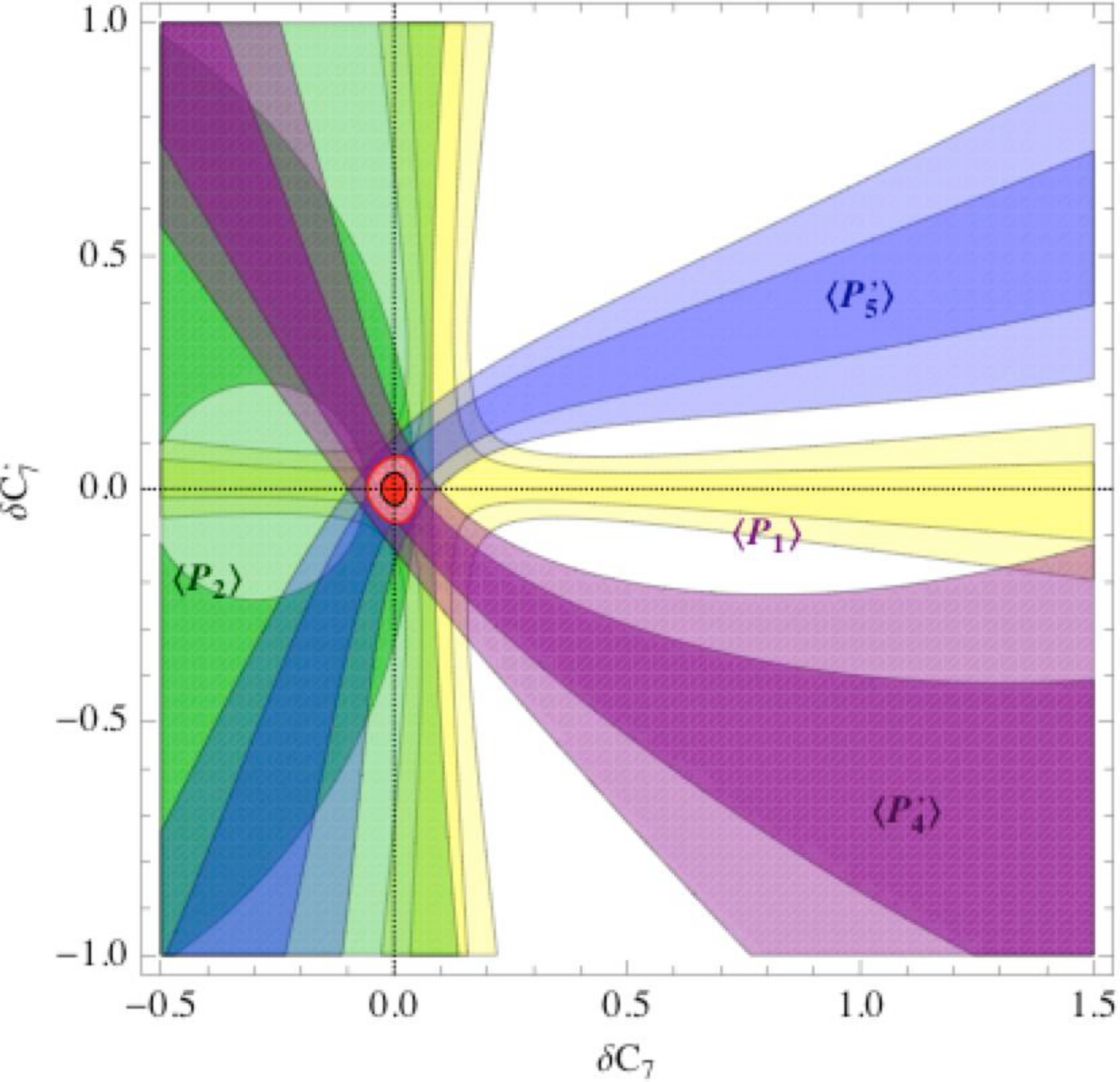}
\caption{Future scenario for constraints from $P_i$ observables.}
\label{JVfut}
\end{figure}

To finish, we comment on the prospects for constraints in the $C_7-C_7'$ plane from $\langle P_i\rangle_{\rm bin}$ observables. We consider the situation in which $\langle P_1\rangle_{[2,4.3]}$, $\langle P_2\rangle_{[2,4.3]}$, $\langle P'_4\rangle_{[2,4.3]}$ and $\langle P'_5\rangle_{[2,4.3]}$ are measured, with central values equal to their SM predictions and experimental uncertainties of $\sigma_{exp}=0.10$ (note that this experimental precision is feasible soon). In Fig.~\ref{JVfut} we show the 68.3\% and 95.5\% C.L. combined constraints on $C_7,C_7'$ from these observables. Comparing this plot with Fig.~\ref{JVcons} we can see that the observables $\langle P_i\rangle$ will play a very important role in the future.

\begin{theacknowledgments}
 
It is a pleasure to thank the organizers of the conferences QCD@Work'2012 and FLASY'12 for the arrangement of very stimulating workshops. J.V. is supported in part by ICREA-Academia funds and FPA2011-25948.
 
\end{theacknowledgments}

\bibliographystyle{aipproc}   

\end{document}